\documentclass{article}

\usepackage{arxiv}

\usepackage[utf8]{inputenc} 
\usepackage[T1]{fontenc}    
\usepackage{hyperref}       
\usepackage{url}            
\usepackage{booktabs}       
\usepackage{amsfonts}       
\usepackage{nicefrac}       
\usepackage{microtype}      
\usepackage{lipsum}
\usepackage{graphicx}
\usepackage{amsmath}

\title{Optical Reservoir Computing using multiple light scattering for chaotic systems prediction}

\author{
  $^{1,2}$Jonathan Dong, $^{1}$Mushegh Rafayelyan, $^{2}$Florent Krzakala, $\&$ $^1$Sylvain Gigan \\
 $^1$ Laboratoire Kastler Brossel, Sorbonne Université, École Normale Supérieure–Paris Sciences et Lettres (PSL)\\ Research University, CNRS, Collège de France, 24 rue Lhomond, 75005 Paris, France\\
 $^2$ Laboratoire de Physique de l'\'Ecole Normale Sup\'erieure, Universit\'e PSL, CNRS, \\
 Sorbonne Universit\'e, Universit\'e Paris-Diderot, Sorbonne Paris Cit\'e, Paris, France
  }

\begin{document}

\maketitle

\begin{abstract}
Reservoir Computing is a relatively recent computational framework based on a large Recurrent Neural Network with fixed weights. Many physical implementations of Reservoir Computing have been proposed to improve speed and energy efficiency. In this study, we report new advances in Optical Reservoir Computing using multiple light scattering to accelerate the recursive computation of the reservoir states. Two different spatial light modulation technologies, namely, phase or binary amplitude modulations, are compared. Phase modulation is a promising direction already employed in other photonic implementations of Reservoir Computing. Additionally, we report a Digital-Micromirror-based Reservoir Computing at up to 640 Hz, more than double the previously reported frequency using a remotely controlled optical device developed by LightOn, and present new binarization strategies to improve the performance of binarized Reservoir Computing. 
\end{abstract}


\section{Introduction}

%
%
%
%

Optical computing uses the speed and parallelism that light provides in order to process information efficiently \cite{ambs2010optical}. For half a century, considerable efforts have been carried out to develop an all-optical computer, but such a dream has never been realized due to the fierce competition of electronics. Still, optics has found successful applications in long-range communication and specialized computations. The success of Machine Learning has sparked a revival of interest in photonic neuromorphic computing \cite{denz2013optical}. Machine Learning algorithms have recently been applied to a large variety of tasks, ranging from image recognition to Natural Language Processing, and  the neural networks used in Machine Learning are well-suited for optical computing. In particular, an optical phenomenon, called light scattering, has successfully been applied to Echo-State Networks \cite{dong2018scaling}, a class of networks part of the more general framework of Reservoir Computing.

In complex media, light does not propagate in a straight line, as refractive index inhomogeneities alter the direction of propagation. For example, turbulent airflows in the atmosphere and turbid biological tissues distort images and prevent the acquisition of a well-resolved image. This effect has been intensively studied in astronomy and biological microscopy, to image through or inside these complex media. Adaptive optics techniques can compensate the effect of scattering using wavefront shaping \cite{tyson2010principles, maurer2011spatial}. For even stronger distortion, when light propagates through a one-hundred-micron-thick layer of white paint for instance and gets scattered by small particles at random positions, direct imaging is no longer possible as the image has been scrambled by multiple scattering events. This process results in a complex interference figure, called speckle pattern. In this regime, it is possible to describe the scattering process by a linear multiplication with a dense matrix, called the Transmission Matrix (TM), which is characteristic of the medium and has been shown to be random \cite{popoff2010measuring}.
In other words, after the complex medium, the information of the incident electric field has been mixed but not lost. Instead of trying to reverse the detrimental effect of scattering to image through complex media, we want to explore the use of this mixing of information as a resource for computation. 
Controlling the incident field by using Spatial Light Modulators (SLMs), 
we can efficiently perform high-dimensional matrix multiplication and obtain an optical computing strategy that is potentially competitive with state-of-the-art electronic devices for various Machine Learning applications.

Random matrix multiplications are present in Echo-State Networks \cite{jaeger2001echo, jaeger2004harnessing} and Liquid-State Machines \cite{maass2002real}. They are Recurrent Neural Networks with fixed generic interconnection weights between neurons. Training these networks is considerably simpler compared to other Recurrent Neural Networks with tunable interconnection weights. Only the final linear layer is trained to predict the correct output by solving a simple linear regression problem. This class of networks represents a promising numerical tool for understanding and predicting temporal datasets, and they have found successful applications such as speech recognition \cite{triefenbach2010phoneme}, robot motor control \cite{antonelo2008event}, or financial forecasting \cite{lukovsevivcius2012reservoir}. 

Reservoir Computing (RC) unifies and generalizes the previous approach, it transforms the previous recurrent network with fixed weights into a generic reservoir \cite{verstraeten2007experimental}. The reservoir is not necessarily described by a neural network, any system with rich and stable dynamics can be leveraged for Reservoir Computing. This opens up the possibility to use unconventional substrates \cite{tanaka2019recent}, where a physical reservoir receives an external excitation from an input and predictions are read out from an observation of the reservoir state. In the last decade, there has been a strong interest to find an efficient physical implementation, from dedicated electronic boards like Field-Programmable Gate Arrays \cite{antonik2015fpga} and memristive devices \cite{donahue2015design}, to photonic circuits \cite{van2017advances, bueno2018reinforcement} and carbon nanotubes \cite{dale2016evolving}. 

\begin{figure*}[!t]
\centering
\includegraphics[width=\linewidth]{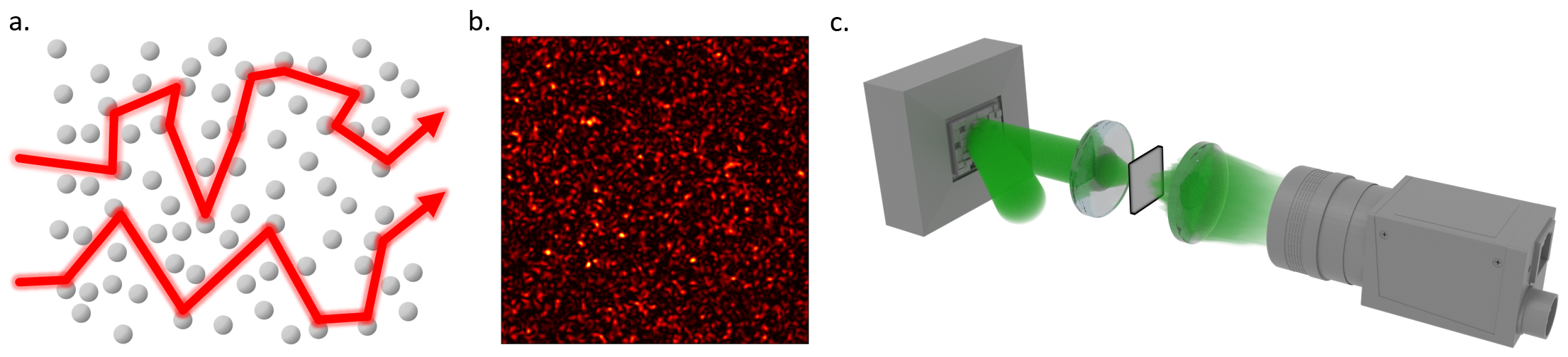}
\caption{Multiple light scattering for optical random projections. 
(a) When light propagates in a complex medium, it undergoes many scattering events, modifying its direction of propagation. 
(b) This scattering process results in a complex interference figure, called a speckle pattern, after the complex medium. 
(c) Experimental setup for optical random projections. The light from a laser (Ti:Sapphire laser at 800 nm, MaiTai, Spectra Physics) is modulated by a Spatial Light Modulator (Meadowlark 512$\times$512 LCoS Reflective SLM), sent in a random scattering medium (a diffuser 0.5 mm thick), and collected on a camera (Allied Vision, Manta G-046). The speckle pattern collected on the camera is a random projection of the incident SLM image.}
\label{light_scattering}
\end{figure*}

In this study, we use optical scattering to implement Reservoir Computing algorithms for chaotic time series prediction. Here, only the iterative computation of the reservoir state is accelerated optically; these successive reservoir iterations represent the most computationally demanding operation, while we still perform on a conventional computer the other steps related to the final linear layer. This optical computing strategy can outperform electronic implementations by two orders of magnitude in speed and scale to very large dimensions that we cannot reach in electronics due to memory limitations. In this paper, we improve over the previous optical realization of binary Echo-State Networks presented in \cite{dong2018scaling} thanks to the introduction of binary encodings to increase binary RC performance, as well as introducing a new non-binary RC algorithm that uses spatial phase shaping of light instead of the previously reported on-or-off intensity modulation.

The choice of the Spatial Light Modulator (SLM) technology to convert digital information into an analog electric field is critical. We investigate here two approaches, either the use of a Digital Micromirror Device (DMD) or an SLM based on Liquid Crystal on Silicon technology (LCoS). The former, with DMDs, can only display binary images and this constraint needs to be addressed efficiently to preserve the performance of the associated Reservoir Computing algorithm. On the other hand, LCoS-SLMs can imprint 8-bit phase images on the optical electric field, which facilitates this digital to analog conversion, but they are relatively slow compared to Digital Micromirror Devices (few tens of Hertz versus up to 20kHz). 

We tested here these optical networks on chaotic system prediction, with the Mackey-Glass equations. As the reservoir is a complex dynamical system itself, Reservoir Computing is intrinsically linked with chaos and it can be trained as a non-linear predictor of chaotic time series. For instance \cite{pathak2018model} reports state-of-the-art performance on the chaotic Kuramoto-Sivashinsky equation, while other applications in chaos synchronisation and cryptography have been demonstrated \cite{antonik2018using}, revealing the intricate link between Reservoir Computing and chaos.

The main body of this article is organized as follows. We introduce the concepts of multiple light scattering in Section 2 and Reservoir Computing in Section 3. We detail the optical implementation in Section 4. Section 5 and 6 present the experimental results with LCoS-SLM and DMD respectively, introducing new binary encoding strategies in the last section. 

\section{Light scattering}

\subsection{General background}

When light encounters refractive index inhomogeneities, it gets scattered and its direction of propagation is modified. Fog and white paint are typical volumetric scattering samples, with  water droplets and titanium dioxyde pigments being the respective scattering particles. When light propagates through thick scattering samples, the number of scattering events is tremendously high. It is impossible to precisely describe all the light propagation in the medium, and at the exit of the scattering medium, one typically observes a speckle figure. This image results from a complex interference process where all the different scattering paths are recombined. Thanks to the large number of scattering events at random positions, this speckle image is seemingly random and its statistical properties are well characterized \cite{goodman2007speckle}. It represents a signature of the particular disordered medium and, for a given incident field, will be different from one scattering sample to another. 

Originally developed for astronomy and biological imaging, wavefront shaping techniques can control light and even perform imaging, in this multiple scattering regime. For example, it is possible to focus light after a scattering medium, by modulating many incident modes to generate a constructive interference at a specific point \cite{vellekoop2007focusing}. This has given rise to a wealth of applications in imaging and beyond \cite{rotter2017light}.

\subsection{The Transmission Matrix of complex scattering media}

One particular object of interest for our current study is the Transmission Matrix (TM) \cite{popoff2010measuring}. Light propagation, even in the multiple scattering regime, remains a linear process: the output over a set of detectors can be described as the product between the incident electric field on a set of input modes, and the TM. This matrix thus characterizes light propagation between input and output modes. In practice, due to the finite size of our optical devices and not measuring back-reflected light, we only measure a sub-part of a theoretical complete Scattering Matrix that would fully describe light propagation in the disordered medium, both in transmission and reflection \cite{rotter2017light}. 

We place a Spatial Light Modulator (SLM) and a camera at the two sides of a multiple scattering medium, to collect transmitted light through the scattering sample. An SLM is a device that is commonly used for wavefront shaping, it imprints a given image as a phase or amplitude modulation on an incident light field. The SLM displays an input image on the coherent light coming from a laser and this modulated electric field will propagate through the complex material to form a speckle figure on the camera. 

This speckle, although seemingly random, depends on the input image on the SLM. This link is made explicit in the Transmission Matrix, that describes the linear relationship between the input electric field defined by the SLM image and the output speckle pattern. The camera records the intensity of the output electric field, its image is thus given by:
\begin{equation}
    b = |H a|^2
\end{equation}
where $a$ is the electric field on the plane of the SLM and $H$ is the TM describing the propagation of light from the SLM to the camera, through the scattering medium. 

The TM can be measured experimentally and it has been proven to be i.i.d random in the multiple scattering regime \cite{popoff2010measuring}.  The matrix dimensions are determined by the number of controlled input and output modes, which are fixed by the resolution of the associated optical devices. Nowadays, common Spatial Light Modulators (SLMs) and cameras typically contain a few million pixels, thus a Transmission Matrix can reach the gigantic size larger than $10^6 \times 10^6$. We cannot possibly hope to measure such a large matrix, as it would take a prohibitive time, and it would be impossible to store it in the memory of a computer. However, we can leverage the very large dimensionality of the TM without measuring it: by displaying a vector on the SLM as an input before the disordered medium, we effectively multiply this vector by the very large TM and record the modulus of the result on the camera. Therefore we want to use the statistical properties of multiple scattering \cite{goodman2007speckle} for optical computing. 

\subsection{Spatial Light Modulators for optical computing}

To have an efficient optical computing strategy, light modulation needs to be performed efficiently. There are several technologies available, as they have been developed for the display industry. Three important criteria are the speed of the modulation, the total number of pixels on the SLM, and the encoding depth of the SLM image. We document here Reservoir Computing implementations using two kinds of Spatial Light Modulators: Digital Micromirror Devices (DMDs) and SLMs based on the Liquid-Crystal on Silicon technology (LCoS-SLMs). 


A Digital Micromirror Device is made of an array of micromirrors that can be set in two positions. In one state, light will be sent to the scattering medium, in the other, light will be deflected towards a beam blocker and won't contribute to the output speckle. Hence, using a DMD, we can display binary 0/1 images. Developed by Texas Instruments since the end of the 80s, it is now a mature technology as it is present in most videoprojectors nowadays. The speed of the DMD is their main advantage for optical computing, as they can reach tens of kHz. Mirrors can be switched on and off very quickly to produce greyscale for consumer displays, as the eye integrates in time the intensity of every pixel of an image. This method cannot be used for speckle patterns because it comes from interferences that depend on the electric field and not the intensity. Averaging the intensity of different random speckle patterns in time decreases the contrast of the final speckle pattern. 

In contrast, phase-only LC-SLMs use twisted nematic liquid crystals to modulate the phase of an incident electric field. The orientation of the liquid crystals in a small region can be modulated by an imposed electric field, and this induced change will modulate the index of reflection, thus the phase of the reflected light accordingly. Organized in arrays of pixels, these Spatial Light Modulators are able to imprint a digital phase image coming from the computer. Their speed is limited by the response time of the liquid crystal system, and the fastest version available today reaches about 500 Hz (without taking into account data transfer). 

\subsection{Apart\'e on random projections}

Propagation through a multiply light scattering medium results in applying a random matrix multiplication on the original input vector, an operation known as a random projection.  As a matter of fact, random projections have been studied extensively in computer science as a convenient way to transform the dimension of a set of vectors while preserving its structure. The Johnson-Lindenstrauss lemma \cite{johnson1984extensions} states that pairwise distances are preserved as long as the dimension after the random projection is logarithmic in the number of elements in the original set, regardless of the original dimension of the vectors. Hence, random projections have been used in data compression and to reduce computational complexity in Randomized Linear Algebra \cite{mahoney2011randomized, woodruff2014sketching}. Several machine learning algorithms also use random projections, such as Random Features and Reservoir Computing, and they are particularly suited for optical computing thanks to their robustness against analog noise \cite{saade2016random, dong2018scaling}. In a nutshell, a random projection is a convenient generic operation that preserves distances between vectors. Thus, even if the speckle appears to be random without structure, it still contains information about the vector displayed on the SLM.

\section{Reservoir Computing}

\begin{figure*}[!t]
\centering
\includegraphics[width=\linewidth]{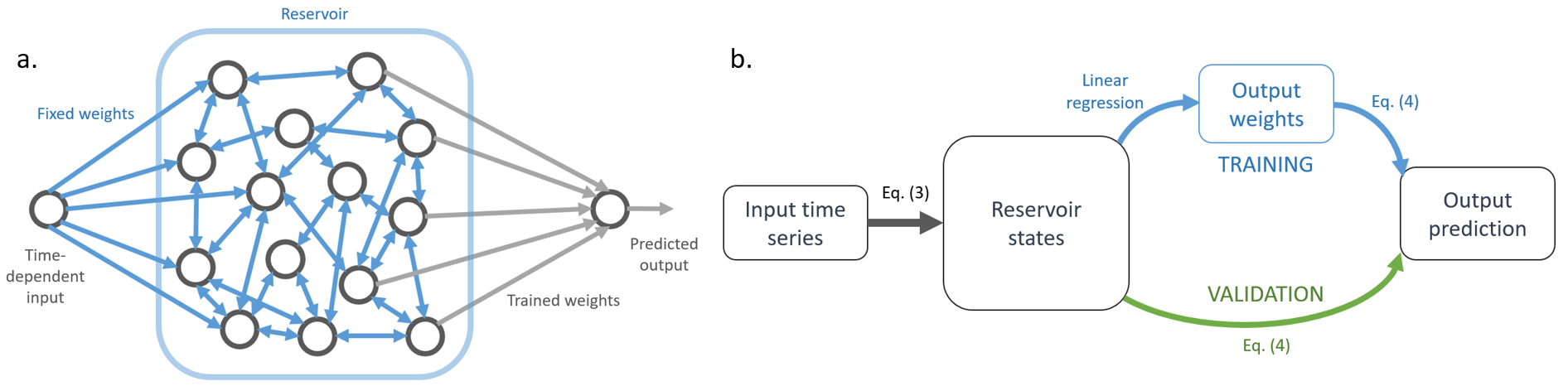}
\caption{Reservoir Computing principle. (a) A Recurrent Neural Network with fixed weights is used as a reservoir to encode information from a time-dependent input. It can be replaced by another dynamical system. The output weights are trained with a linear regression. (b) Reservoir Computing flowchart. The successive reservoir states are first computed, then used for training or validation.}
\label{fig_sim}
\end{figure*}

\subsection{Background}

Machine learning is a powerful tool to learn patterns and make inferences in complex problems based on a large number of examples. It relies on models with tunable parameters, for example Neural Networks with interconnection weights between neurons tuned to perform a particular task. In supervised learning, these weights are trained using a large number of examples, which consist in pairs of input data with the desired outputs. Thanks to increasing computational power and a large amount of available data to process, they have achieved state-of-the-art performance on very diverse tasks, such as image recognition \cite{krizhevsky2012imagenet}, Natural Language Processing, or recommender systems \cite{bell2007lessons} to name a few. Today machine learning is a blooming field, and there is a number of other machine learning approaches beyond neural networks such as kernel methods and decision trees. Computational efficiency is a major research direction, as there is a strong need to scale down the heavy machine learning computations in smaller devices.

Recurrent Neural Networks are notoriously hard to train \cite{pascanu2013difficulty}. Recurrent connections are a challenge for error back-propagation, which is the method of choice to train neural networks with feed-forward connections such as Convolutional Neural Networks \cite{lecun1998gradient}. Back-propagation through time is possible \cite{werbos1990backpropagation}, but this method faces the problem of local minima, as well as exploding and vanishing gradients \cite{pascanu2013difficulty}. As a possible solution to bypass this training issue, Echo-State Networks (ESNs) \cite{jaeger2001echo, jaeger2004harnessing} are Recurrent Neural Networks with randomly fixed internal weights. Only the output weights are trained for a particular task, reducing the training to a simple linear regression. The number of tunable parameters is thus smaller, but this does not necessarily mean that this Neural Network model is less expressive than fully-tunable Recurrent Neural Networks. It is very easy to increase the number of neurons and it has been proven that large networks can universally approximate any continuous function \cite{grigoryeva2018echo}.

\subsection{Recursive equation}

To operate an Echo-State Network, a time-dependent input is first fed to the network with fixed weights. After the computation of all the ESN states, the output weights are either learned with a training dataset containing the desired outputs, or used to obtain a predicted output on another dataset for validation.  

Let $\{i(t), t = 0, ..., T\} \in \left(\mathbb{R}^d\right)^T$ be an input time series of dimension $d$ and of length $T$. The ESN will be initialized in a random state $x(0) \in \mathbb{R}^n$ and its state at time $t$ will be denoted $x(t)$; $n$ is the dimension of the network, i.e. the number of neurons or reservoir nodes. 
Let $W_{\rm{res}}$ be the internal weight matrix and $W_{\rm{in}}$ the weight matrix between the input and the network. Both weight matrices are random and fixed for Echo-State Networks. The nonlinear activation function of every neuron will be denoted $f$.
The successive ESN states are computed using the following recursive equation:
\begin{equation}
	x(t+1) = f(W_{\rm{in}} i(t) + W_{\rm{res}} x(t))
	\label{esneq}
\end{equation}

In other words, an Echo-State Network is a large set of neurons randomly interconnected, that evolves dynamically driven by an external input. This leads to the more general framework of Reservoir Computing, where the neural network can be replaced by any non-linear dynamical system. An input time series is fed to the reservoir, and the reservoir can be any generic dynamical system. At any time, the current state of the reservoir depends on the previous values of input data, it encodes this information in its state. 

Thanks to this general framework, we introduce three objects that will enable us to tune the dynamics of the reservoir for optical implementations: a leak rate $a$, a random bias vector $b$, and an encoding function $g$ into the ESN equation.
\begin{equation}
	x(t+1) = (1-a) x(t) + a f\big(W_{\rm{in}} g(i(t)) + W_{\rm{res}} g(x(t)) + b\big)
	\label{rceq}
\end{equation}
The leak rate $a$ is an important parameter that controls the speed of the dynamics of the reservoir without changing its long-term stability, the bias parameter $b$ controls the diversity of neurons inside the reservoir states. Finally, the encoding function will be important in the following as images on the Spatial Light Modulator need to be either binary or phase-only. Compared to the implementation of \cite{dong2018scaling} that did not use such a function but enforced contraints directly in the activation function, the network states are real which increases the performance considerably. 

\subsection{Final linear layer}

The output $o(t) \in \mathbb{R}^k$ is computed using a linear combination with weights $W_{\rm{out}} \in \mathbb{R}^{k \times n}$, which can be written as:
\begin{equation}
	o(t) = W_{\rm{out}} x(t)
\end{equation}
The optimal set of output weights is obtained by solving a linear regression problem, which minimizes the following error metric:
\begin{equation}
	E = \frac{1}{k} \sum_t \| \tilde{o}(t) - W_{\rm{out}} x(t) \|^2
	\label{error}
\end{equation}
where $\tilde{o}(t)$ is the target output at time $t$.

Linear regression is a well-studied problem and many libraries already provide an efficient solver. We use the Ridge solver of the scikit-learn library in Python as the ridge regularization is important to avoid overfitting when the number of parameters, which is proportional to the number of neurons, is larger than the number of examples. It correspond to the addition of a term $\alpha \|W_{\rm{out}}\|^2$ in Equation (\ref{error}).

\subsection{Physical implementations of Reservoir Computing}

The flexibility of RC, that can use any generic high-dimensional reservoir for computation, makes it very promising for physical implementations \cite{tanaka2019recent, van2017advances}. Many have been proposed originating from very different research areas, such as optical nano-circuits \cite{vandoorne2014experimental} or carbon nanotubes \cite{dale2016evolving}, one early RC implementation even observed the ripples at the surface of a bucket of water for pattern recognition \cite{fernando2003pattern}. The RC framework is robust against noise and changes in network topology (dense, sparse, and local connections can be used). In principle, various physical systems receive an external time-dependent excitation and follow non-linear dynamics, sending the sequential input into a high-dimensional feature space. To make a prediction, the output is obtained by a linear combination of the observed state of the reservoir. 

Following the recent trend of specialized electronic processors for efficient machine learning \cite{sze2017hardware}, neuromorphic electronic circuits for RC have been developed, based on analog circuits, FPGAs \cite{antonik2015fpga} or memristive devices \cite{donahue2015design}. 

Optical node arrays have also been proposed, where semiconductor optical amplifiers are used to perform the non-linear activation function \cite{vandoorne2014experimental}. Increasing the number of neurons means increasing the number and the density of optical nodes, which is a challenging manufacturing task to address. 

On the other hand, some optical reservoir computers use a single physical node \cite{appeltant2011information, larger2012photonic, paquot2012optoelectronic}. They use temporal multiplexing and are based on a delay-fiber system to generate interconnections between units. They can be operated very fast at the GHz frequency and there is a very active community pursuing this line of research. In this case, a large number of neurons decreases the effective frequency of this time-multiplexed scheme as more nodes need to be sent one by one in the delay system.

Our strategy to increase the dimension of the reservoir is to use free-space optics, combined with high-dimensional cameras and SLMs. The manufacturing challenge of producing cameras and SLMs with a large number of pixels has already been solved thanks to their use in the display industry. By exploiting this commodity, it is possible to implement large-scale Optical Reservoir Computing, interconnection between units being either provided by a Diffractive Optical Element (DOE) \cite{bueno2018reinforcement, antonik2018performance} or a scattering medium \cite{dong2018scaling}. In the first case, connections are typically local and they can be engineered when designing the DOE. In the second case that we study here, the weight matrix is dense and random, which is closer to the original ESN definition. With these devices, we can reach very large sizes but the operating frequency is limited by the SLM, which is typically in the tens or hundreds of Hertz, or more often by the camera. An implementation using an LCoS-SLM in phase modulation to generate the reservoir couplings has also been proposed \cite{pauwels2018towards}.


\subsection{Analysis of the reservoir dynamics}

The reservoir in Reservoir Computing is a tunable dynamical system, that can be put in a stable or chaotic regime depending on a few parameters controlling the dynamics. Empirically, best performance in Reservoir Computing happens when the reservoir is “at the edge of chaos” \cite{schrauwen2009computational}, which corresponds to a stable regime close to a chaotic one that exhibits rich but stable dynamics. In practice, the dynamics is controlled empirically by a few parameters and some hyper-parameter search is often used to find a particular dynamical system suited for the task at hand. Note that this procedure is common in Machine Learning as we often need to search for the best model to use for a particular problem by tuning a set of hyper-parameters. 

To describe the complex dynamics of the Reservoir, several properties and quantities of interest have been proposed. The first and most important one, called the Echo-State Property \cite{jaeger2001echo}, describes the stability of the dynamical system. It states that a reservoir needs to forget its initial state after a finite time to be used for RC. In other words, for a given input time series, the observed state of a reservoir after a warm-up phase shall not depend on the method used to initialize the network, which is not related with the task to solve. The reservoir needs to operate in a stable dynamical regime, in contrast to a chaotic regime that would be very sensitive to initial conditions. As a consequence, a chaotic dynamical system cannot be used for Reservoir Computing. Additionally, to ensure that the reservoir state does not depend on the initial conditions, we typically throw away the first states following the initialization, both for training and prediction. 


Two other properties to understand the reservoir dynamics are the separation and approximation properties \cite{maass2002real}. The separation property means that two different input time series need to lead to different final reservoir states. A complex reservoir with rich dynamics is able to encode more information and differentiate a larger number of inputs. The approximation property states that a single input time series perturbed with noise should consistently be mapped to the reservoir state. These properties are useful to characterize in a simple way the high-dimensional dynamics of a reservoir. For instance, it is possible to derive quantitative measures of these properties on experimental Reservoir Computers, and even display the observables introduced in this subsection in a 3D space to visualize the performance of a particular RC implementation \cite{dale2018substrate}. 

In the end, there are four requirements for a dynamical system to be effectively used in RC \cite{tanaka2019recent}: high-dimensionality of the reservoir, non-linearity in the dynamics, the Echo-State Property, and a balance between the separation and approximation properties which depends on the task to solve.



\begin{figure*}[!t]
\centering
\includegraphics[width=.8\linewidth]{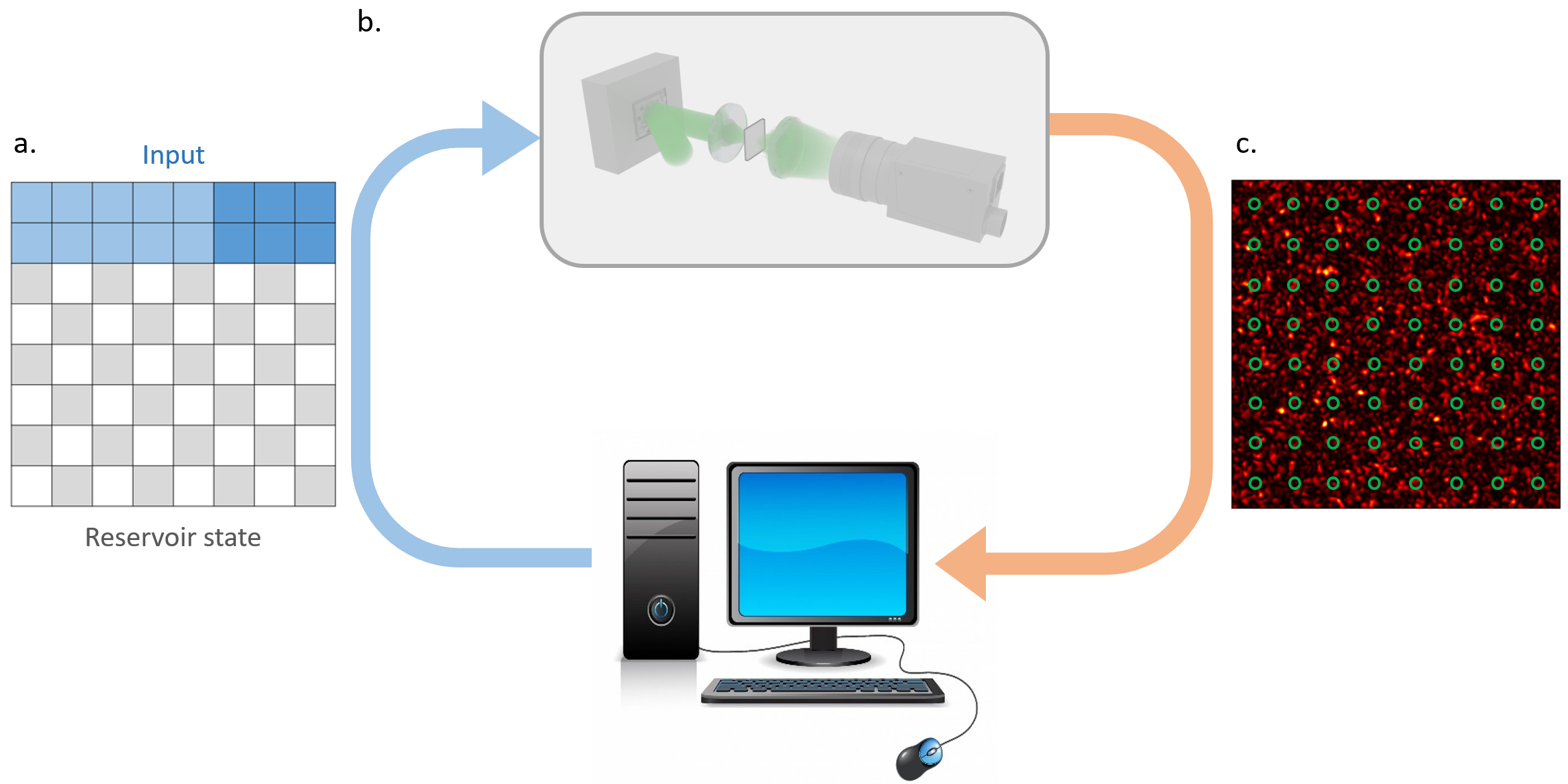}
\caption{Principle of our optical implementation. (a) Scheme of an image sent to the Spatial Light Modulator (DMD or SLM). The current input and reservoir state are encoded and displayed on the Spatial Light Modulator. (b) The optical setup is used to perform the random projection. All the other steps are performed on a computer. (c) Speckle subsampling on the camera image. The next reservoir state is determined by camera values on a grid with spacing larger than the speckle grain size to minimize correlations.}
\end{figure*}

\section{Optical Implementation}

\subsection{General principle}

The experimental setup is depicted in Figure \ref{light_scattering}. The coherent light from a laser is enlarged using a telescope to fill the active area of the SLM, in order to maximize the number of pixels that can modulate the electric field. We use the SLM to display an image sent by the computer, formed from the current reservoir state and input data. The SLM modulates the coherent electric field and reflects it towards a scattering medium, that is thick enough to ensure a complete mixing of the incident modes, so that the Transmission Matrix is dense and not structured. 

The resulting speckle pattern is captured by a camera, determining the next reservoir state, that will be encoded and displayed back on the SLM with the next input. This feedback loop between the camera and the SLM corresponds to one Reservoir Computing iteration and it will be repeated as many times as there are reservoir states to compute. 

The optical implementation with a DMD is developed by LightOn and available for researchers as a cloud service \cite{lighton}, with a very similar implementation as in \cite{saade2016random}. We built a similar optical experiment with an LCoS-SLM instead of a DMD, to investigate the use of phase modulation in Optical Reservoir Computing. The wavelength of the laser we used is at 800 nm (Ti:Sapphire laser, MaiTai, Spectra Physics, operated in continuous-wave mode), we expand the beam to fill the SLM, a Meadowlark 512$\times$512 LCoS Reflective SLM. The modulated electric field is then focused by a 20$\times$ objective with 0.4 numerical aperture on a 0.5 mm thick scattering material. The scattered field is collected by another similar objective and the resulting speckle is captured by a CCD camera (Allied Vision, Manta G-046). The distance from the medium to the camera is adjusted in order to obtain speckles grain size larger than the pixel size. 

\subsection{Encoding information on the Spatial Light Modulator}

At every RC iteration, the input and the reservoir state are concatenated and displayed on the SLM, but due the physical constraints of the SLM, the displayed image needs to be either binary amplitude or phase-only. An encoding operation is introduced at this step, it corresponds to the function $g$ in (\ref{rceq}). In the case of phase encoding, $g$ is simply defined by $g(x) = e^{i \pi x}$. $x$ is between 0 and 1 after normalization and encoded in 8 bits, which correspond to 255 grey levels that are sufficient to avoid any loss of performance due to discretization. On the other hand, the binarization constraint introduced by the DMD is more critical and how to encode a real-valued number will be discussed in Section VI. 

The relative area of the input and the reservoir state is also an important parameter to control. By performing a hyper-parameter search on a noiseless simulation model for the same time-series prediction task, we have seen that the RC implementation gets optimal performance when the input area is around ten times larger than the reservoir area. However, in this case the contribution of the reservoir is very small: one-tenth of the signal after one iteration and approximately one-hundredth after two. Such a small perturbation would be quickly lost due to experimental noise which would be detrimental for the memory capacity of the reservoir, so we chose to fix the reservoir area to be equal to the input area. 

\subsection{Reading the camera image}

\begin{figure*}[!t]
\centering
\includegraphics[width=.8\linewidth]{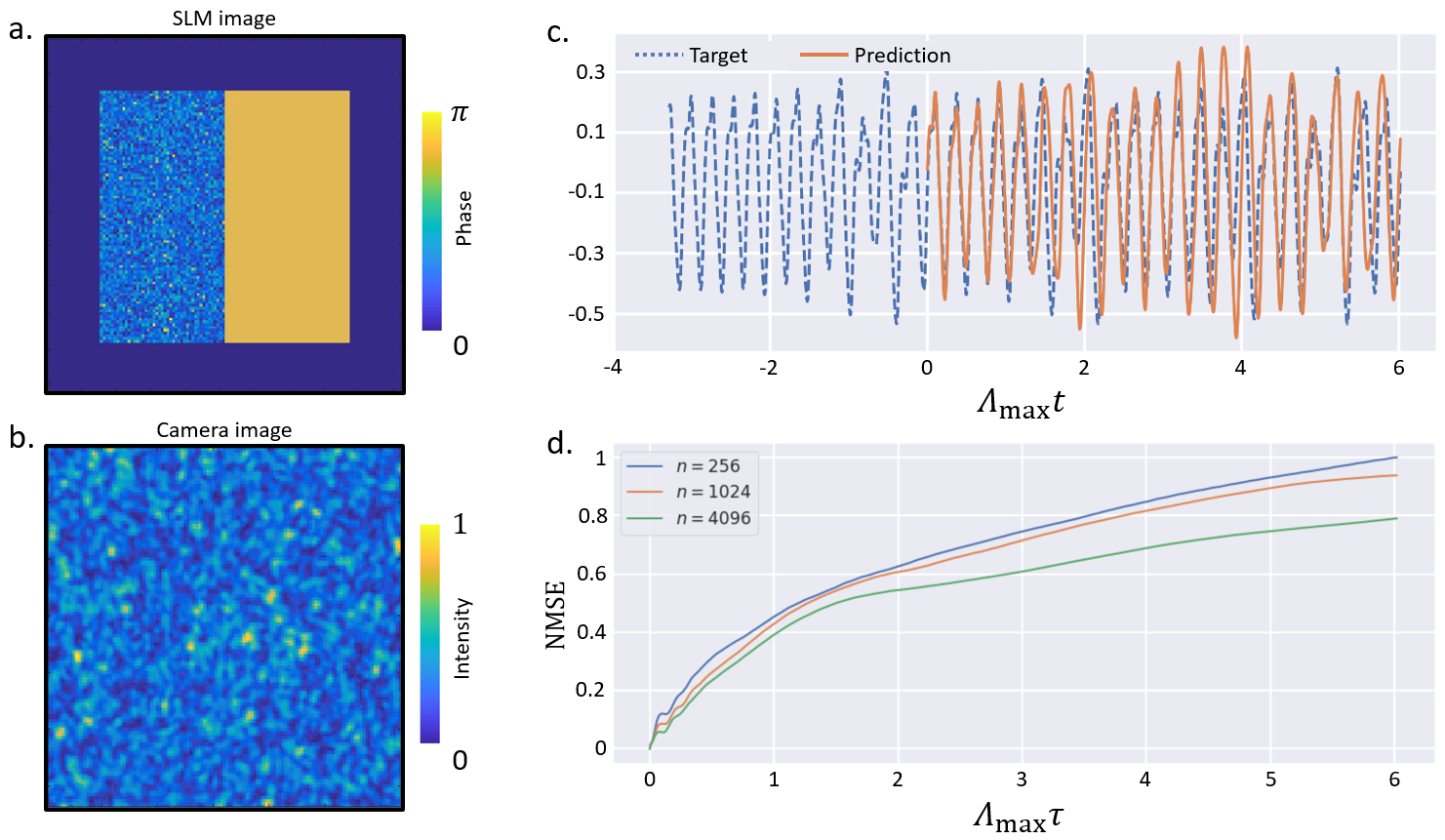}
\caption{Optical implementation results of Reservoir Computing with multiple light scattering and an LCoS-SLM. (a) Encoded phase profile on the SLM. The reservoir state (left) and current time series value (right) are displayed in the central region, while the remaining area is left constant to provide a bias term. Each component of the vector to display is encoded in phase between 0 and pi. (b) Intensity profile captured by the camera during one iteration of Reservoir Computing. (c) Example of prediction on the chaotic Mackey-Glass dataset. The time series is fed to the reservoir until time $t_0=0$, and from the reservoir state at time $t_0$, we predict the future evolution of the time series. The time axis is normalized with the Lyapunov exponent. (d) NMSE for different reservoir sizes.}
\label{slm_fig}
\end{figure*}



After the propagation in the scattering medium, the camera returns a speckle pattern to the computer. This image represents a random projection of the data displayed on the SLM. This random image presents small speckle grains that are a few pixels wide, their size is determined by the diffraction of the finite numerical aperture of the optical system. We choose a sampling grid larger than the speckle grain size in order to remove this local range correlation and make sure that the interconnection matrix is fully random. This property can be been tested by computing the distribution of singular values of the Transmission Matrix, that should follow the prediction of Random Matrix Theory and is modified when local correlations are present in the matrix \cite{popoff2010measuring}. 

The intensity values of the speckle pattern follows an exponential distribution, as they are the absolute value square of a complex Gaussian random variable \cite{goodman2007speckle}. Its mean depends on the laser power, the thickness of the scattering medium, and the exposure time of the camera. We empirically observed better performance when using the square root of the intensity, i.e. the modulus of the electric field, instead of the raw intensity measured by the camera, probably because this operation regularizes the reservoir state distribution for the subsequent linear regression.



\subsection{Chaotic time series prediction}

Mackey and Glass proposed in 1977 their famous equation to model physiological feedback systems:
\begin{equation}
	\frac{du(t)}{dt} = \beta \frac{u(t-\tau)}{1+u(t-\tau)^n} - \gamma u(t)
\end{equation}
The first term corresponds to a delayed response of the system, which tends to 0 as $u$ tends to either 0 or infinity to keep the model realistic, while the second term can be interpreted as a classical decay with rate $\gamma$. This time-delay differential equation, in appearance simple, displays chaotic behavior for certain ranges of parameters, for example $\beta = 0.2$, $\gamma = 0.1$, $\tau = 17$, $n = 10$ as in \cite{jaeger2004harnessing}. The maximal Lyapunov exponent in this case is $\Lambda_{\rm{max}}=0.006$. In general, the Lyapunov exponent gives a measure for the total predictability of a system, it characterizes quantitatively the rate of separation of infinitesimally close trajectories in dynamical system, namely, the minimum amount of the time for which trajectories are diverging by a factor of $e$. This chaotic time series is one of the standard models used to test Reservoir Computing algorithms \cite{jaeger2001echo}.


\section{Mackey-Glass time series prediction: Reservoir Computing with a phase SLM}

\subsection{Experimental parameters}

Two different Mackey-Glass series with $T = 2000$ time-steps are randomly generated, for training and testing the RC algorithm with the SLM. The prediction task consists in two phases. In the first phase, the algorithm constructs the reservoir from the given test data $u(t_0), u(t_0 - 1), \ldots$, then it predicts the next $\tau$ values, $u(t_0+1), \ldots, u(t_0+\tau)$, using already constructed reservoir. The larger the prediction time $\tau$, the harder the associated prediction task becomes. Besides, the larger the reservoir size, the better the prediction performance. Therefore, the tasks with large prediction times require large reservoir sizes. The RC algorithm with SLM is tested for three typical reservoir sizes, $n = 256, 1024, 4096$, and for multiple prediction time periods $\tau = 1,2, \ldots, 1000$. 

The SLM screen is split into three regions of equal areas, corresponding to the input, the reservoir state, and the bias. The bias region stays constant during the whole RC iterations and it provides a reference speckle on the camera that increases the diversity inside the reservoir states. Figure \ref{slm_fig}a presents an example of the encoded phase profile on the SLM and the captured intensity speckle pattern on the camera during one iteration of RC. The input information at every iteration is a single value, since the Mackey-Glass time series is one-dimensional. Therefore, the same input value is encoded onto multiple SLM pixels to respect the equal importance ratios between input, reservoir and bias regions. The same macro-pixel strategy is applied for the reservoir information when the size of the reservoir is smaller than the number of available pixels on the SLM.

The SLM splitting ratios with other important parameters that affect the performance of the RC algorithm, included the ridge regression parameters and the leak rate, are obtained by grid search using the scikit-learn library \cite{scikit-learn}. The leak rate is set to 0.3, we forget the first 100 initial reservoir states in regression, and the training regularization parameter is set to 10000.

This high value of the regularization parameter helps to compensate experimental noise, especially short-term fluctuations coming from detector noise, SLM flickering or mechanical vibrations. There are also long-term deviations coming from the decorrelation of the scattering medium to take into account that also affect the system performance.

\subsection{Experimental results}

Figure \ref{slm_fig}c shows one of the successful prediction results. The dashed line is the test time series fed to the reservoir until $t_0 = 0$. Afterwords, the algorithm has been switched into prediction mode (solid line). The temporal axis is normalized by the maximal Lyapunov exponent $\Lambda_{\rm{max}}$.

As a benchmark of the prediction performance we use the $\rm{NMSE}$ (Normalized Mean Square Error) for $N$ values of $t_0$ defined by:
\begin{equation}
    \text{NMSE} = \frac{E}{N \sigma^2},
    \label{NMSE}
\end{equation}
where $E$ is the squared error on the prediction task presented in Equation (\ref{error}), normalized by $\sigma^2$ the variance of the Mackey-Glass time series. Consequently, the smaller the $\rm{NMSE}$ value, the better the prediction performance is. In order to obtain smoother $\rm{NMSE}$ curves, we calculated its average over ten independent experiments with nine hundred test time series for each. The performance curves we obtain with the SLM optical implementation for different reservoir sizes $n_{\rm{res}} = 256,\ 1024,\ 4096$ are collected in Figure \ref{slm_fig}d. As one can see, the larger the reservoir size, the better the algorithm performance. The small oscillations of the $\rm{NMSE}$ from short prediction times originate from the oscillations of the Mackey-Glass time series, which are faster than the Lyapunov time. For longer prediction times, the task becomes harder which explains why the performance reaches a kind of plateau. The task becomes exponentially harder since the Mackey-Glass time series is chaotic and the Lyapunov exponent defines the typical time scale for chaotic divergence. It should converge to 1 for much longer time series as the reservoir only outputs the mean value of Mackey-Glass.

The speed of the algorithm is determined by the working frequency of the SLM and camera, as well as the data transfer speed between the computer and the optical devices. In our particular case, the setup is performing around 70 iterations per second. The main advantage of our optical implementation is its scalability. Namely, we can easily reach large reservoir sizes as the speed of the implementation almost does not depend on the size of the reservoir \cite{dong2018scaling}. 


\section{Quantized Reservoir Computing: binary encoding for DMD implementation}

\subsection{Quantization in Reservoir Computing}

\begin{figure}[!t]
\centering
\includegraphics[width=0.5\linewidth]{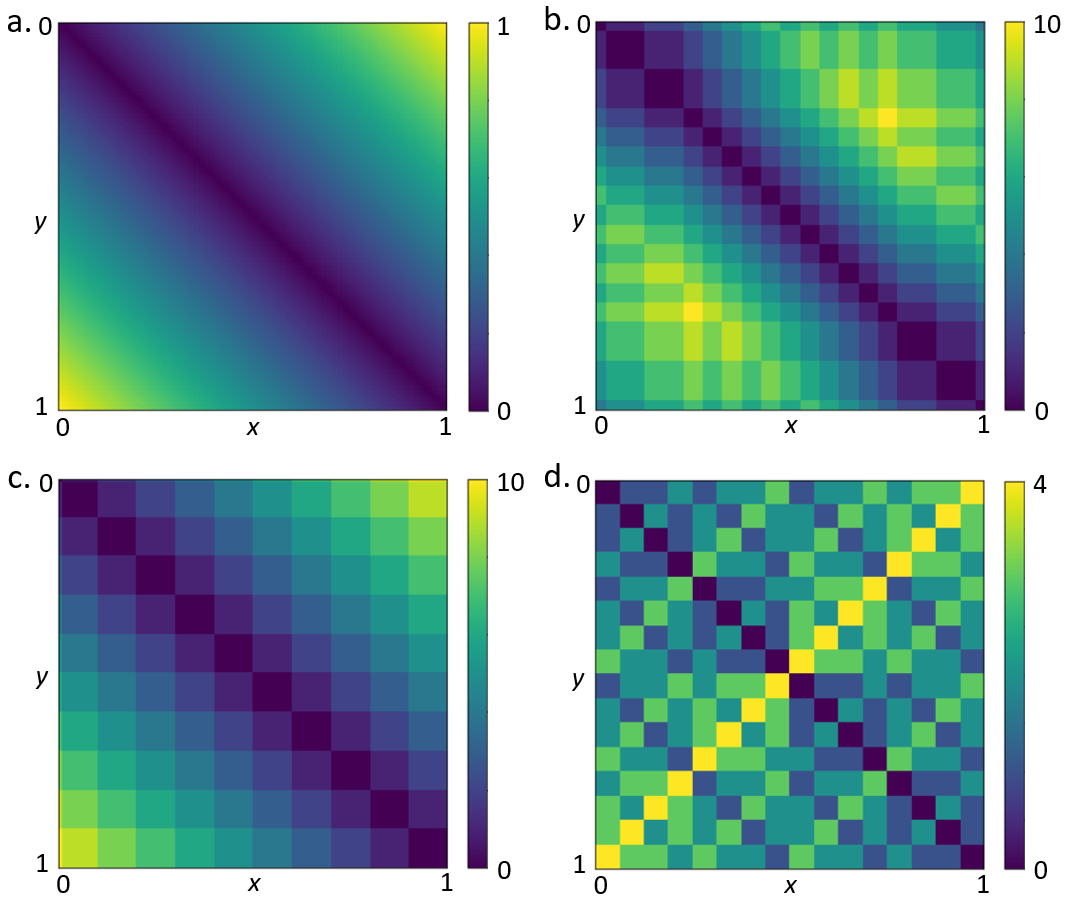}
\caption{Distance matrices to compare encoding strategies. (a) Distance matrix $\|g(x) - g(y)\|$ for $x$ and $y$ between 0 and 1, with no encoding here $g(x) = x$ (for control). (b) Distance matrix for basket encoding (\ref{basket_enc}) with $n_{\rm{bin}} = 10$. (c) Distance matrix for threshold encoding (\ref{threshold_enc}) with $n_{\rm{bin}}=10$. (d) Distance matrix for base-2 encoding with $n_{\rm{bin}}=4$.}
\label{encoding}
\end{figure}

DMDs are based on the actuation of micromirrors to modulate the incident wavefront, they can operate at much faster rates than LCoS-SLMs but only a binary modulation is possible. In Reservoir Computing, this introduces a necessary binarization step in the dynamics of the Reservoir, which induces a discretization and loss of information. This quantization task has also been studied for other applications in Machine Learning \cite{courbariaux2016binarized}, as electronic devices are generally more efficient to perform summations than multiplications. They show that neuronal activations or weights can be binarized without notably decreasing the performance of the model. Thus, the architecture choice and high number of neurons are able to compensate the loss introduced by quantization. 

It has been shown that binary Echo-State Networks can successfully learn to predict the chaotic Mackey-Glass time series \cite{dong2018scaling}. In this first implementation, the activation function was a simple threshold function on the intensity, an operation which loses a lot of information. However, in Reservoir Computing, the binarization step still has a significant impact on the reservoir dynamics and performances are generally worse than models without binarization. One explanation of this degradation is that binarization considerably modifies the stability of the reservoir dynamics, as every quantization threshold is a discontinuity and very similar inputs can lead to well-separated states. Hence the usual theoretical tools based on Lipschitz continuity to prove the Echo-State Property do not hold for a quantized activation function \cite{jaeger2001echo}. The dynamics of binary models, and in particular the transition between stability and chaos, has been studied in \cite{schrauwen2009computational}.


\subsection{Binarization strategy}

\begin{figure*}[!t]
\centering
\includegraphics[width=.8\linewidth]{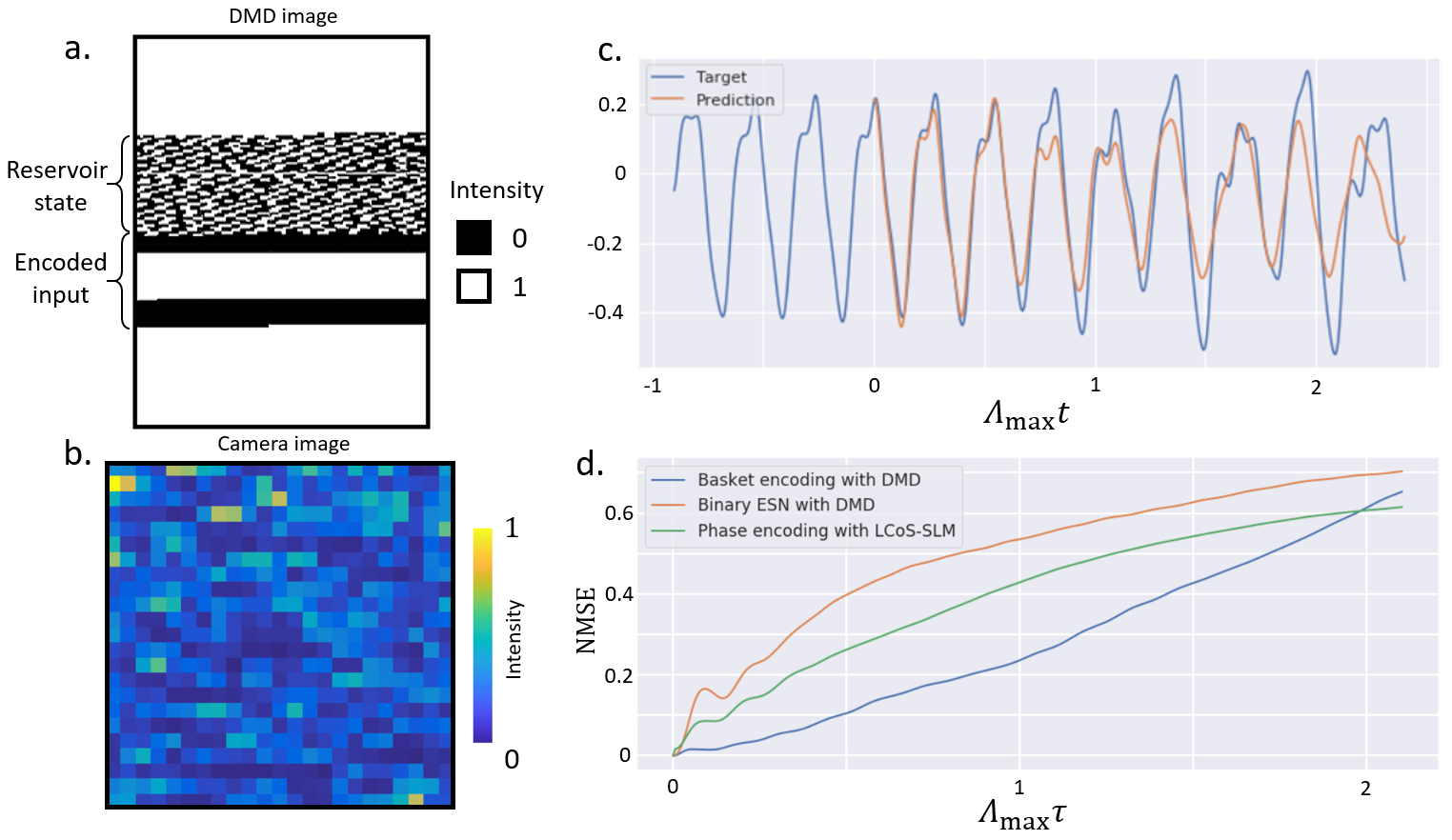}
\caption{Optical implementation results of Reservoir Computing with multiple light scattering and a DMD. (a) Binary amplitude image displayed on the DMD image during a RC iteration. The encoded reservoir state and input are displayed on approximately half of the DMD area, while the rest provides a constant random bias. (b) Camera image defining the reservoir state at the next iteration. Each component of the real-valued vector to display on the DMD is encoded in 10 binary values using basket encoding. (c) Example of prediction on the chaotic Mackey-Glass dataset. The time series is fed to the reservoir until time $t_0=0$, and from the reservoir state at time $t_0$, we predict the future  evolution of the time series. The time axis is normalized with the Lyapunov exponent. (d) NMSE for different realizations of Optical Reservoir Computing: DMD implementation with basket encoding, DMD implementation with binary ESN  \cite{dong2018scaling}, and LCoS-SLM with phase encoding. Reservoir sizes are 512, 5120, and 1024 respectively.}
\label{dmd_fig}
\end{figure*}



To minimize the loss of information due to binarization, we use here the camera image to define the reservoir state. This change improves the RC performance considerably. Here the activation function $f$ is simply a modulus square operator, without a threshold operation. The binarization is performed in another step, it corresponds to the introduction of the encoding function $g$ in \ref{rceq} that operates componentwise, i.e. on each pixel in parallel. It does not necessarily modify the dynamics of the system, but we keep the reservoir state as rich as possible by storing the camera image instead of a quantized version. 

Thanks to the introduction of this encoding function, we can also change the dimensionality of the encoding and encode a reservoir activation on several DMD pixels. With this strategy, the function $g$ encodes the given value of one camera pixel on $n_{\rm{bin}}$ binary DMD pixels. Higher values of $n_{\rm{bin}}$ increase the regularity of the encoding function, and make the RC dynamics smoother. 

Such an encoding also requires more DMD pixels to display the reservoir state. Hence, an efficient binarization scheme needs to balance between two constraints: a limited dimension expansion, to allow very large reservoir to be displayed on the DMD, and sufficient regularity and precision of the encoding.

There are several possibilities to define the encoding function $g$. As a separable function, i.e. operating componentwise, it is uniquely defined by the image of a single real number. After normalization of the number to encode, we assume that this number lies between 0 and 1. Intuitively, an efficient binarization strategy should send two close values on similar binary encodings and distant values on dissimilar encodings. Hence, the characteristics of an encoding are well-represented by its distance matrix, defined by the set of $\|g(x) - g(y)\|$ for all pairs $(x,y) \in [0;1]^2$. These distances should be small close to the diagonal, where $x$ and $y$ are similar, and increase further away from the diagonal. Kernel methods also study how distances and scalar products are transformed by non-linear embeddings, thus there is a close link between these binary embeddings and kernel methods.

Inspired by the Random Binning Features developed by Rahimi and Recht \cite{rahimi2008random}, we propose here an encoding function $g$, called basket encoding, where each component $g_i(x)$ for $i = 1, \ldots, n_{\rm{bin}}$ is defined by:
\begin{equation}
    g_i(x) = \left\{ \begin{array}{lr}
    1, & \text{if } x \in \left[c_i-s, c_i+s\right] \\
    0, & \text{else}
    \end{array}\right.
    \label{basket_enc}
\end{equation}
where the centers and size of the bins are defined by $c_i = \frac{2i-1}{2n_{\rm{bin}}}$ and $s = \frac{2 \lfloor n_{\rm{bin}} / 2 \rfloor - 1}{4n_{\rm{bin}}}$ respectively. These values are chosen to obtain a larger number of different binary encodings while keeping a regular distance matrix. 

Another binarization strategy, previously used in \cite{dong2018scaling} and referred to as threshold encoding, is to encode the input by using a set of uniformly-spaced thresholds:
\begin{equation}
    g'_i(x) = \left\{ \begin{array}{lr}
    1, & \text{if } x > t_i \\
    0, & \text{else}
    \end{array}\right.
    \label{threshold_enc}
\end{equation}
where the thresholds are defined by $t_i = \frac{i}{n_{\rm{bin}}}$ for $i = 1, \ldots, n_{\rm{bin}}$. This scheme is also regular and can be used for RC, but the number of different binary encodings is smaller in this case (10 possible binary encodings for threshold encoding compared to 15 for basket encoding). As a result, for a fixed binary encoding dimension $n_{\rm{bin}}$, it will contain less information about the original real-value.

Fig. \ref{binary_sim} presents the distance matrices for these three different binarization strategies. We observe that both basket encoding and threshold encoding are well-behaved, as distances close to the diagonal, which correspond to close original real values, are small while the distance increases smoothly as the original values get further apart. Additionally, the basket encoding provides more precision for a given bit depth, here $n_{\rm{bin}} = 10$.

We also present the distance matrix using the representation in base 2. This binary encoding is the one used in the memory of a computer, it is the most compact one as $n_{\rm{bin}}$ represent $2^{n_{\rm{bin}}}$ different numbers. However, the computer implicitly differentiates bits according to their position, from the most-significant to the least-significant bit; in RC all bits should have the same importance. For example, 7 and 8 (or $\frac{7}{16}$ or $\frac{8}{16}$ after normalization) are very close but their binary encodings in base 2, 0111 and 1000, are very different.

\subsection{Cloud implementation}

We use an optical device developed by LightOn, available on the cloud. It is based on the same principle as the LCoS SLM presented previously, but uses a DMD for binary wavefront modulation. This device performs the optical random projections using multiple light scattering presented in Section 2 and is open for researchers to use.

In every optical RC iteration, we need to use the camera image to compute the next DMD image to display. However, the LightOn device is optimized to send DMD images by batches to maximize the speed of data transfer. We therefore compute batches of reservoirs in parallel driven by different time series, between 500 and 3'000. In the end, as we set the time series length to 800, we typically obtain tens of thousand of examples for training and testing. 

Figure \ref{dmd_fig}ab presents the images sent to the DMD and the new state of the reservoir captured on the camera. On the DMD, we encode the current input value and reservoir state in a $1140 \times 912$ binary image. We display the input using basket encoding and 1000 pixels on an area representing approximately one fourth of the total DMD surface. The reservoir state of dimension $n_{\rm{res}}$ is encoded in ten times more binary DMD pixels using basket encoding. The reservoir also occupies one fourth of the DMD, which leaves one half of the DMD as a bias. 

The leak rate is set to 0.2, we forget the 50 initial reservoir states, and the regularization parameter is set to 0.1 for training. 

\subsection{Prediction results}

We present in Figure \ref{dmd_fig}d the $\rm{NMSE}$ of binary Reservoir Computing, binary Echo-State Network and the phase Reservoir Computing obtained with an LCoS-SLM. Reservoir sizes are set at 512 for binary Reservoir Computing, and 5120 for binary ESN to obtain a fair comparison as the basket encoding expands the binary dimension 10 times. We observe that thanks to the basket encoding, this new binarized version of Reservoir Computing is performing better than the previous binary ESN implementation. Additionally, we see that this experimental binary Reservoir Computing is performing better than the other experimental implementation of Reservoir Computing based on phase modulation. The gap in performance between DMD and LCoS-SLM implementations is probably due to a difference in stability and SNR of the optical devices, which is higher for the one developed by LightOn. On the other hand, binary Echo-State Networks do not perform as well due to the binarization operation.

\begin{figure}[!t]
\centering
\includegraphics[width=.5\linewidth]{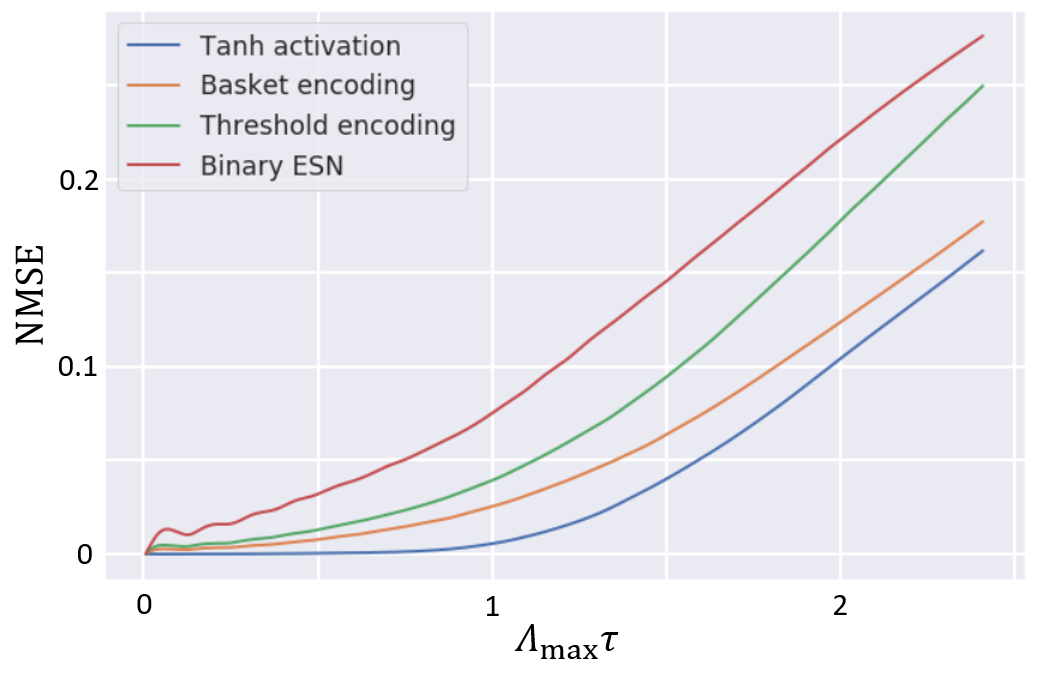}
\caption{NMSE of Mackey-Glass prediction with RC using different encoding strategies: (a) tanh activation, (b) basket encoding defined in (\ref{basket_enc}), (c) threshold encoding defined in (\ref{threshold_enc}), and (d) binary ESN of \cite{dong2018scaling}. Each curve is an average of 5 realizations (numerical simulation), reservoir sizes is 512 for all cases with binary encoding dimension equal to 10 for (b) and (c).}
\label{binary_sim}
\end{figure}

Figure \ref{binary_sim} presents  simulation results to compare different encoding strategies. We see that the basket encoding performs better than the other two binary strategies, threshold encoding and the binary ESN without encoding. This proves that it is important to have a regular and precise encoding for binary Reservoir Computing. 

The speed of the DMD depends on the batch size, as the LightOn device is more efficient when one sends a large number of images to be displayed in a burst. Hence, we achieve 640 Hz with a batch size of 3'000 and 285 Hz with a batch size of 500. We are up to 2 times faster than the previously-reported speed in \cite{dong2018scaling} thanks to hardware optimization. An even faster implementation should be possible by sending directly the camera image to be displayed on the DMD with dedicated electronics, thus avoiding the computer in the RC loop to reduce latency. 

\section{Conclusion}

This study presents how multiple light scattering can be harnessed for Reservoir Computing. Thanks to the complex interference that results from the scattering of light inside a complex medium, we can generate the successive states of a reservoir responding to a given input time series. Optical Reservoir Computing algorithms have been demonstrated for chaotic time series prediction, based on two different SLM technologies. On the one hand, we use an LCoS SLM to perform Reservoir Computing based on phase modulation of the electric field. On the other hand, the DMD implementation represents a promising solution for efficient Reservoir Computing thanks to its high working frequency, but it can only display a binary image. Encoding and binarization strategies have been proposed to make binary Reservoir Computing perform on par with real-valued networks. 

This optical computing strategy can also be applied to other machine learning tasks where random projections prove useful, be it for their distance conservation properties or to emulate any fully-connected neural network with randomly fixed weights. Optical random projections have been demonstrated for instance in image recognition \cite{saade2016random} and change-point detection \cite{keriven2018newma}.

\section*{Acknowledgements}

We would like to thank Antoine Boniface for lending his experimental setup to perform the SLM experiment and a careful review of the manuscript, Claudio Moretti for his help in designing figures, LightOn for providing OPU access and Charles Brossollet for technical support with the OPU. This material is based upon work supported by the Defense Advanced Research Projects Agency (DARPA) under Agreement No. HR00111890042. Sylvain Gigan and Jonathan Dong also acknowledge partial support from H2020 European Research Council (ERC) (Grant 724473).

\bibliography{refs}
\bibliographystyle{ieeetr}

\end{document}